\documentstyle[12pt]{article}
\begin{document}
\begin{flushright}
IHEP 95-87\\
hep-ph/9507242\\
\end{flushright}
\vspace*{4mm}
\begin{center}
{\large\bf Hadronic production of baryons, containing two heavy quarks}\\
{\rm A.V.Berezhnoy, V.V.Kiselev, A.K.Likhoded \\
{\it Institute for High Energy Physics, Protvino 142284, Russia}\\
E-mail: likhoded@mx.ihep.su }
\end{center}
\begin{abstract}
In the framework of QCD perturbation theory, total and differential
cross sections of the $\Xi_{bc}'$, $\Xi_{bc}^{(*)}$ and $\Xi_{cc}^{(*)}$
baryons production in gluon collisions are calculated in the leading order over
$\alpha_s$ for the doubly heavy ($b c$) and ($ c c$) diquarks. At
both small and large transverse momenta of the baryons, a use of the mechanism
of the heavy quark fragmentation into the heavy diquark is shown to
underestimate the cross section values in comparison with the exact numerical
calculations of complete set of diagrams. The expected in Tevatron experiments
yield of baryons with two heavy quarks is evaluated as
$(1.3\pm 0.3) \cdot 10^5 \,\, bcq$-baryons and
$(1.6\pm 0.3) \cdot 10^4 \,\, ccq$-baryons at $p_T>5$ GeV and
$|y|<1$ of the baryon momentum and rapidity cuts, with account for the
antiparticle yields.
\end{abstract}
\section{Introduction}
One of new directions in the heavy quark physics is related with a description
of baryons with two heavy quarks \cite{1}. In a large heavy quark mass limit,
a typical size of a $Q_1 Q_2$-diquark configuration is much less than the
characteristic radius of light hadrons $r_{Q_1 Q_2}\ll \Lambda^{-1}$.
For light quarks, the heavy-heavy diquark object looks like a heavy antiquark
at low virtualities, so that one can relate the $Q_1 Q_2 q$-baryon
characteristics with the properties of a meson with a single heavy quark
\cite{2}. The form factors of the doubly heavy baryons are straightforwardly
related with those of heavy light mesons \cite{3}, described by the
Isgur-Wise functions, up to the accuracy of a probability for the transition
of the doubly heavy diquark to another one. The latter point of view to the
$Q_1 Q_2$ diquark allows one also to formulate some rules, describing the
process of the $Q_1 Q_2 q$ baryon production \cite{4}.

First, the process under consideration can be represented as the process of a
hard production of the $Q_1 Q_2$ diquark, that further has a hadronization
into the $Q_1 Q_2 q$ baryon. The hadronic production of the diquark with a
mixed flavor is calculated under the analogy with the $Q_1\bar Q_2$ heavy
quarkonium production, on the basis of computations of the fourth
$\alpha_s$-order diagrams in the QCD perturbation theory. Second, the
nonperturbative soft part of the matrix element describes the quark binding in
the $Q_1 Q_2$ diquark  and it is given by the diquark wave function at the
origin.

At present, the basis for the current estimates of the $Q_1 Q_2 q$ baryon
production cross section is the consideration of the $Q_1 Q_2$ diquark
production cross sections at large transverse momenta via the fragmentation
mechanism \cite{4}. The fragmentation function of the heavy  quark
$Q_1\rightarrow (Q_1 Q_2)+\bar Q_2$ is taken in the same form as the function
of fragmentation into the heavy quarkonium $Q_1\bar Q_2$ \cite{5,6}
with the same quantum numbers over the Lorentz group. This approach is quite
excusive and valid in the high energy $e^+e^-$-annihilation, where the
fragmentational mechanism does practically dominate \cite{6}, but the
situation is more complex for the hadronic production \cite{7}.
The complication is caused by that a number of the leading order diagrams,
describing the hadronic production of doubly heavy diquark, is much
greater than the number of diagrams in the $e^+e^-$-interaction. Moreover,
the former diagrams do not allow one completely to interpret them in terms of
the fragmentational mechanism. The fragmentation is not exactly defined
in the hadronic production. As was shown \cite{7}, the diagrams of
the recombinational type give a dominant contribution even at quite large
transverse momenta. So, a simple formula, corresponding to the convolution of
the differential cross section of the heavy quark production with the
function of the heavy quark fragmentation into the $Q_1 \bar Q_2$ quarkonium
or $Q_1 Q_2$ diquark, gives only a cross section evaluation in the order of
magnitude. To obtain an exact result is necessary to take into account all
36 diagrams of the fourth order over $\alpha_s$.

In what follows, we present the results of exact calculations for the
$bcq$ and $ccq$ baryons yields in the framework of the computations for the
heavy quark pair production in the fourth order diagrams of QCD. The difference
between the given approach and the fragmentational one is connected with that
the contribution of the complete set of diagrams is calculated with no
neglection of a part of them. For the states with a various quantum numbers,
the performed calculations result in the yield ratios, different from
those of in the approximate approach of the fragmentation.

\section{Calculation technique}

The calculation technique, applied in the present paper, is analogous to that
of the hadronic production of $B_c$ mesons \cite{6}.
The only difference is due to the binding of two heavy quarks ($Q_1$ and $Q_2$)
in contrast to the binding of a heavy quark with a heavy antiquark.

We suppose that the binding energy in the diquark is much less than the masses
of quarks, composing the diquark, and hence, the quarks are on the mass shells.
Therefore, the quark four-momenta are related with the diquark momentum $P$
in the following way
\begin{equation}
p_1=\frac{m_1}{M}P\;,\;\; \quad p_2=\frac{m_2}{M}P,\label{1}
\end{equation}
where $M=m_1+m_2$ is the diquark mass, $m_{1,2}$ are the quark masses.

In the given approach, the diquark production can be described by the 36
leading order Feynman diagrams, corresponding to the production of four
free quarks in the way of combining of two quarks into the colour antitriplet
diquark with the given quantum numbers over the Lorentz group.
The latter procedure is performed by means of the projection operators
\begin{equation}
\frac{1}{\sqrt{2}}\{ \bar u_1(+)\bar u_2(-) - \bar u_1(-)\bar u_2(+) \}
\end{equation}
for the scalar state of diquark (the corresponding baryon
is denoted as $\Xi_{12}'(J =1/2)$~),
\begin{equation}
\begin{array}{c}
\bar u_1(+)\bar u_2(+), \\
\frac{1}{\sqrt{2}}\{ \bar u_1(+)\bar u_2(-) + \bar u_1(-)\bar u_2(+) \},\\
\bar u_1(-)\bar u_2(-)
\end{array}
\end{equation}
for the vector state of diquark (the baryons are denoted as $\Xi_{12}(J =1/2)$
and $\Xi_{12}^*(J =3/2)$~).

To produce the quarks, composing the diquark, in the $\bar 3_c$-state, one has
to introduce the colour wave function as $\varepsilon_{ijk}/\sqrt{2}$,
into the diquark production vertex, so that $i=1,2,3$ is the colour index of
the first quark, $j$ is that of the second one, and $k$ is the colour index of
the diquark.

The diquark production amplitude $A$ is expressed via the amplitude
$A'$ for the free quark production in kinematics (\ref{1}) as
\begin{equation}
A=\frac{\sqrt{2M}}{\sqrt{2m_1}\sqrt{2m_2}}\frac{R(0)}{\sqrt{4\pi}}
\sum_{h_1,h_2,i,j} P(h_1,h_2)
A'(h_1,h_2,i,j)\frac{\varepsilon_{ijk}}{\sqrt{2}},
\end{equation}
where $h_1,h_2$ are the spiralities of corresponding quarks,
$i,j$ are its colour indices,
$R(0)$ is the diquark radial wave function at the origin,
and the $P(h_1,h_2)$ operators have the following explicit form
($H=h_1+h_2$)
\begin{equation}
P(h_1,h_2)=\frac{1}{\sqrt{2}} (-1)^{h_1-\frac{1}{2}}\delta_{H0}
\end{equation}
for the scalar state, and
\begin{equation}
P(h_1,h_2)=|H|+ \frac{1}{2}\delta_{H0}
\end{equation}
for the vector one.

In the numerical calculations, giving results, which will be discussed in the
next section, we suppose the following values of parameters
\begin{equation}
\begin{array}{l}
\alpha_s=0.2,  \\
m_b=4.9\; {\rm GeV}, \\
m_c=1.7\; {\rm GeV}, \\
R_{bc(1S)}(0)=0.714\; {\rm GeV}^{3/2}, \\
R_{cc(1S)}(0)=0.263\; {\rm GeV}^{3/2},
\end{array}
\end{equation}
where the $R_{bc}(0)$ value has been calculated by means of a numerical
solution of the Schr\"odinger equation with the Martin potential \cite{8},
multiplied by the 1/2 factor, caused by the colour antitriplet state of quarks
instead of singlet one, and the $R_{cc}(0)$ value is taken from ref.\cite{4}
for the sake of convenience of the result comparison.

To calculate the production cross section of the diquarks, composed of two
$c$-quarks, one has to account for their identity. One can easily find, that
the antisymmetrization over the identical fermions leads to the scalar
diquark production amplitude, equal to zero, and it results in that the
amplitude of the vector $cc$-diquark production can be obtained by the
substitution of equal masses in the amplitude of the vector $bc$-diquark
production with the account for the 1/2 factor, following from the identity of
quarks and antiquarks.

We assume, that the produced diquark has  the fragmentation into the baryon,
practically carrying away a total diquark momentum, with  the unit probability.
Therefore, discussing the results, we will talk on the differential cross
sections of $\Xi_{bc}',\  \Xi_{bc}^{(*)}$ and $\Xi_{cc}^{(*)}$ baryons
production, since we suppose that the latters negligibly small deviate from
the differential cross sections of diquarks.

\section{Discussion}

The total energy dependence of the gluonic production cross sections of
$\Xi_{bc}'$ ($\circ$) and $\Xi_{bc}^{(*)}$ ($\bullet$) baryons is shown
on Fig.1 and in Tab.1. To compare, the predictions of the fragmentational
mechanism for  $\Xi_{bc}^{(*)}$ (solid line) and $\Xi_{bc}'$ (dashed line) are
also presented. One can see from the figure, that the fragmentational
production mechanism, assuming the validity of factorization in the cross
section at $ M^2/s\ll 1$ and at large transverse momenta via the formula
\begin{equation}
\frac{d\sigma_{gg\rightarrow\Xi_{bc}'(\Xi_{bc}^{(*)}) \bar b \bar c}}
{dz}=\sigma_{gg \rightarrow b \bar b } \cdot
D_{b \rightarrow \Xi_{bc}'(\Xi_{bc}^{(*)})}(z),
\end{equation}
with $z=2|\vec P|/\sqrt{s}$, does not work at low gluon energies, where it
overestimates the cross section, because of the incorrect evaluation of the
phase space, and it is not valid also at large energies, where the predictions
of the fragmentational mechanism are essentially less that the exact results.
So, the fragmentational values underestimate the
$\Xi_{bc}^{(*)}$ and $\Xi_{bc}'$ cross sections in 9 and 4 times,
respectively, at $\sqrt{\hat s}=100$ GeV. When the fragmentational predictions
give the ratio $\sigma_{\Xi_{bc}^{(*)}}/ \sigma_{\Xi_{bc}'}\simeq 1.4$,
the exact perturbative calculations result in
$\sigma_{\Xi_{bc}^{(*)}}/ \sigma_{\Xi_{bc}'}\simeq 3.2$
even at $\sqrt{\hat s}=100$ GeV.

The agreement with the fragmentational production at $\sqrt{\hat s}=100$ GeV
is poor even at large transverse momenta of the baryon, as one can see from the
distributions over $p_T$ for the $\Xi_{bc}^{(*)}$ and $\Xi_{bc}'$ production,
shown on Fig.2 in comparison with the predictions of the fragmentational
mechanism. Note, that in contrast to the doubly heavy baryon production,
the exact perturbative calculations of the gluonic production of
$B_c(B_c^*)$ mesons with $p_T > 35$ GeV at $\sqrt{\hat s}=100$ GeV
agree with the fragmentational predictions. For the baryon production,
a visible deviation is observed up to the largest values of $p_T$.

The differential cross section $d\sigma /d p_T$ of the
$\Xi_{bc}'(\Xi_{bc}^{(*)})$ production in $p \bar p$ interactions at
$\sqrt{s}=1.8$ TeV is presented on Fig.3 in comparison with the
fragmentational predictions. In the $B_c$ meson production, the
difference between the exact and fragmentational approaches was slightly
hidden due to the convolution with the hadron structure functions. In the
process of baryon production, this is not the case, and the invalidity
of the fragmentational approach explicitly follows from the form of the
distribution under consideration.

The calculation results on the $\Xi_{cc}^{(*)}$ production point out that
the deviation between the exact perturbative and fragmentational values is also
essential as in the $\Xi_{bc}^{(*)}$ production.
This disagreement can be noted on Fig.4 and in Tab.2, where the dependence
of the $\Xi_{cc}^{(*)}$ production cross section on the energy of the gluon
interactions is presented, as well as on Fig.5, where the differential
cross section $d\sigma /d p_T$ of the $\Xi_{cc}^{(*)}$ production in the
gluon interactions is shown at $\sqrt{\hat s}=100$ GeV in comparison with
the fragmentational mechanism predictions. From the latter figure, one can see
that even at $p_T > 40$ GeV, the exact result slightly overestimates the
fragmentation.

The differential cross section $d\sigma /d p_T$ of the $\Xi_{cc}^{(*)}$
production in $p \bar p$ interactions at the energy $\sqrt{s}=1.8$ TeV
is shown on Fig.6 in comparison with the fragmentational estimate. One can see,
that at the reasonable values of $p_T$, the fragmentational result
is approximately 3 times less than the exact perturbative one.

At the chosen values of parameters and with the account for the cuts over
the transverse momentum and rapidity of the baryons ($p_T>5$ GeV and $|y|<1$),
the production cross section of the $1S$-wave $bcq$-baryons and its
antiparticles is evaluated as $\sigma_{bcq} \simeq 1 $ nb, and the total cross
section of the $1S$-wave $ccq$-baryon production with the account for
antiparticles is equal to $\sigma_{ccq} \simeq 0.13 $ nb. After the expected
end
of Run Ib at Tevatron with the integral luminosity $100\div 150\;
\mbox{pb}^{-1}$, one has the yields of $1.0\div 1.5 \cdot 10^5$ of
$bcq$-baryons and $1.3\div 1.9 \cdot 10^4$ of $ccq$-baryons.

Thus, we have shown, that the calculations in the leading order approximation
of the QCD perturbation theory for the gluonic production of the doubly heavy
diquarks, having the hadronization into the baryons, lead to the essential
discrepancy in the differential as well as total cross sections of the
baryon production in comparison with the predictions of the mechanism of
the heavy quark fragmentation into the diquark.

\begin{table}[p]
\begin{center}
\caption{The dependence of the gluonic production cross sections for
the $\Xi_{bc}'$ and $\Xi_{bc}^{(*)}$ baryons on the total energy.
(The error on last digits stands in brackets.)}
\begin{tabular}{||c|c|c||}
\hline
     &             &           \\
$\sqrt{\hat s}$, GeV  & $\sigma_{\Xi_{bc}'}$, pb &
$\sigma_{\Xi_{bc}^{(*)}}$, pb  \\
     &             &           \\
\hline
15.  &  0.4395(9)  & 1.091(3)  \\
20.  &  1.487(3)   & 5.96(2)   \\
30.  &  2.06(1)    & 8.47(4)   \\
40.  &  2.043(18)  & 7.95(6)   \\
60.  &  1.63(3)    & 5.87(9)   \\
80.  &  1.25(4)    & 4.32(11)   \\
100. &  0.98(5)    & 3.31(13)  \\
\hline
\end{tabular}
\end{center}
\end{table}

\begin{table}[p]
\begin{center}
\caption{The dependence of the gluonic production cross sections for
the $\Xi_{cc}^{(*)}$ baryons on the total energy.
(The error on last digits stands in brackets.)}
\begin{tabular}{||c|c||}
\hline
     &             \\
$\sqrt{\hat s}$, GeV  & $\sigma_{\Xi_{cc}^{(*)}}$, pb  \\
     &             \\
\hline
15.  &  3.18(2)   \\
20.  &  3.26(3)   \\
40.  &  1.97(4)   \\
60.  &  1.23(5)   \\
80.  &  0.85(7)   \\
100. &  0.68(5)   \\
\hline
\end{tabular}
\end{center}
\end{table}

\newpage
\section*{Figure captions}
\begin{itemize}
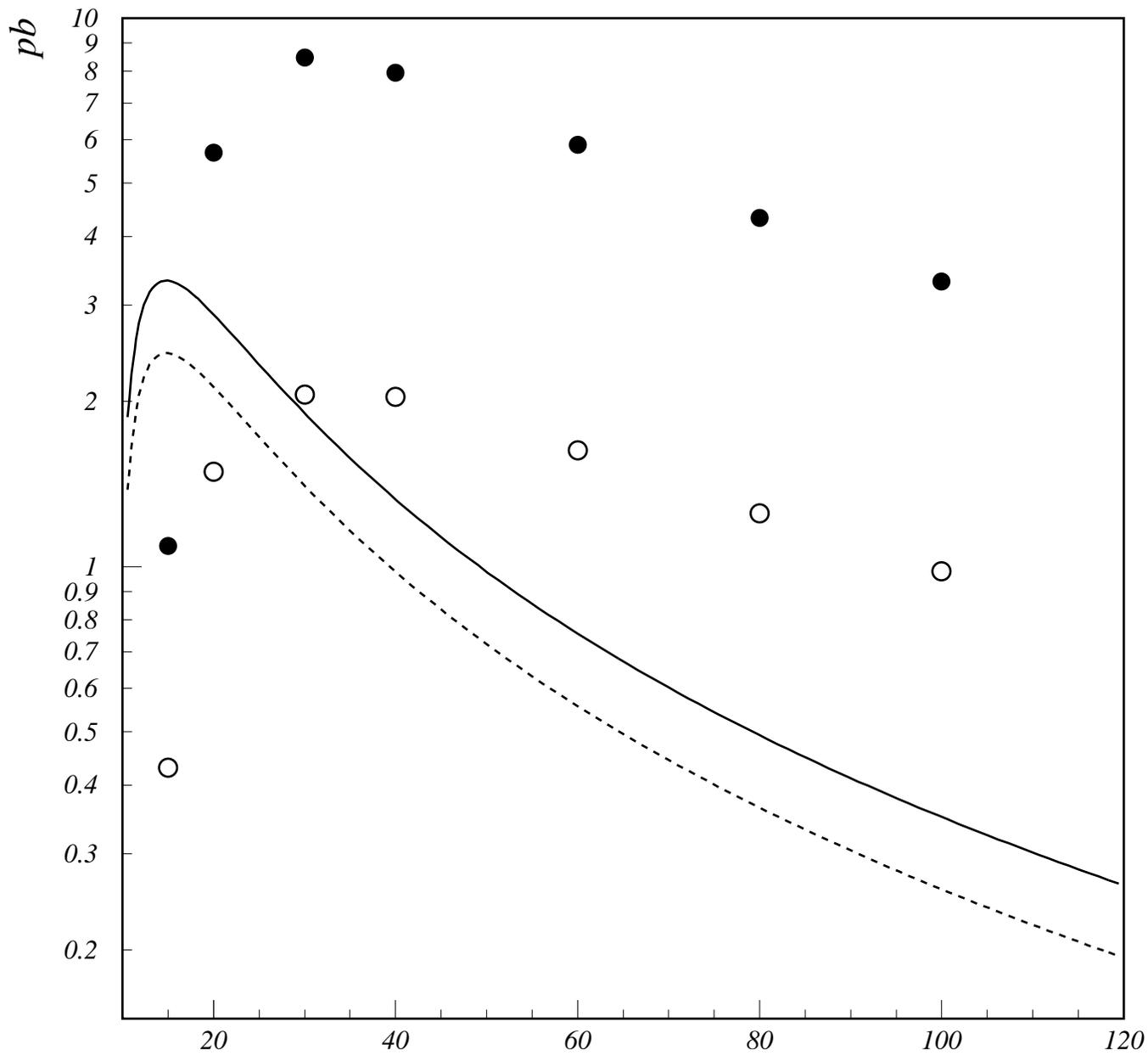
\item[Fig. 1.] The gluonic production cross sections
of $\Xi_{bc}'$ ($\circ$) and $\Xi_{bc}^{(*)}$ ($\bullet$) in comparison with
the predictions of the fragmentational mechanism for
$\Xi_{bc}'$ (dashed line) and $\Xi_{bc}^{(*)}$ (solid line).

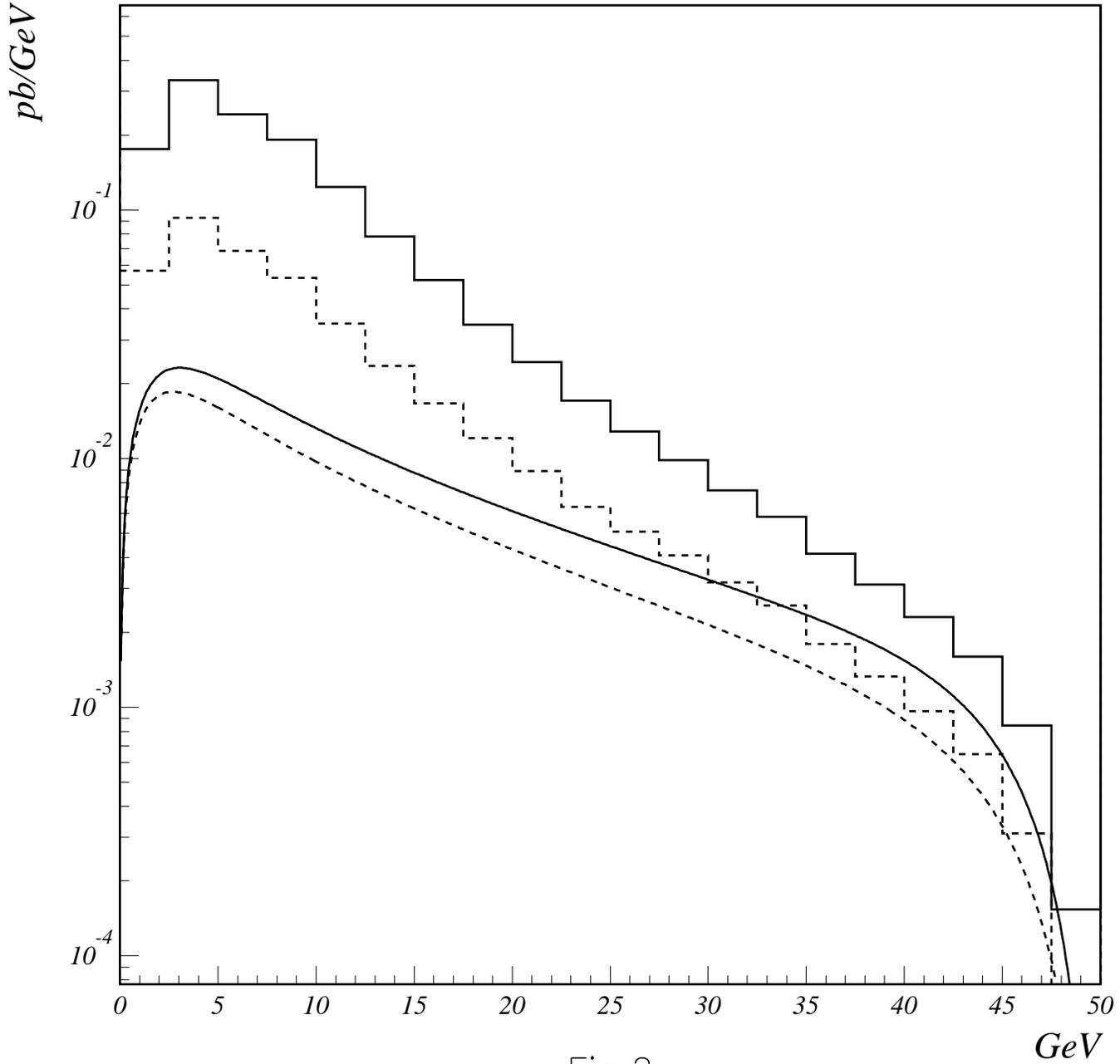
\item[Fig. 2.] The distributions over the transverse momentum in the gluonic
production of $\Xi_{bc}'(\Xi_{bc}^{(*)})$ in comparison with the fragmentation
result at the interaction energy 100 GeV.
Here and in what follows, the solid and dashed lines correspond to the
$\Xi_{bc}^{(*)}$ and $\Xi_{bc}'$ production, respectively, and
the exact perturbative and fragmentational results are shown as
histograms and smooth curves, respectively.

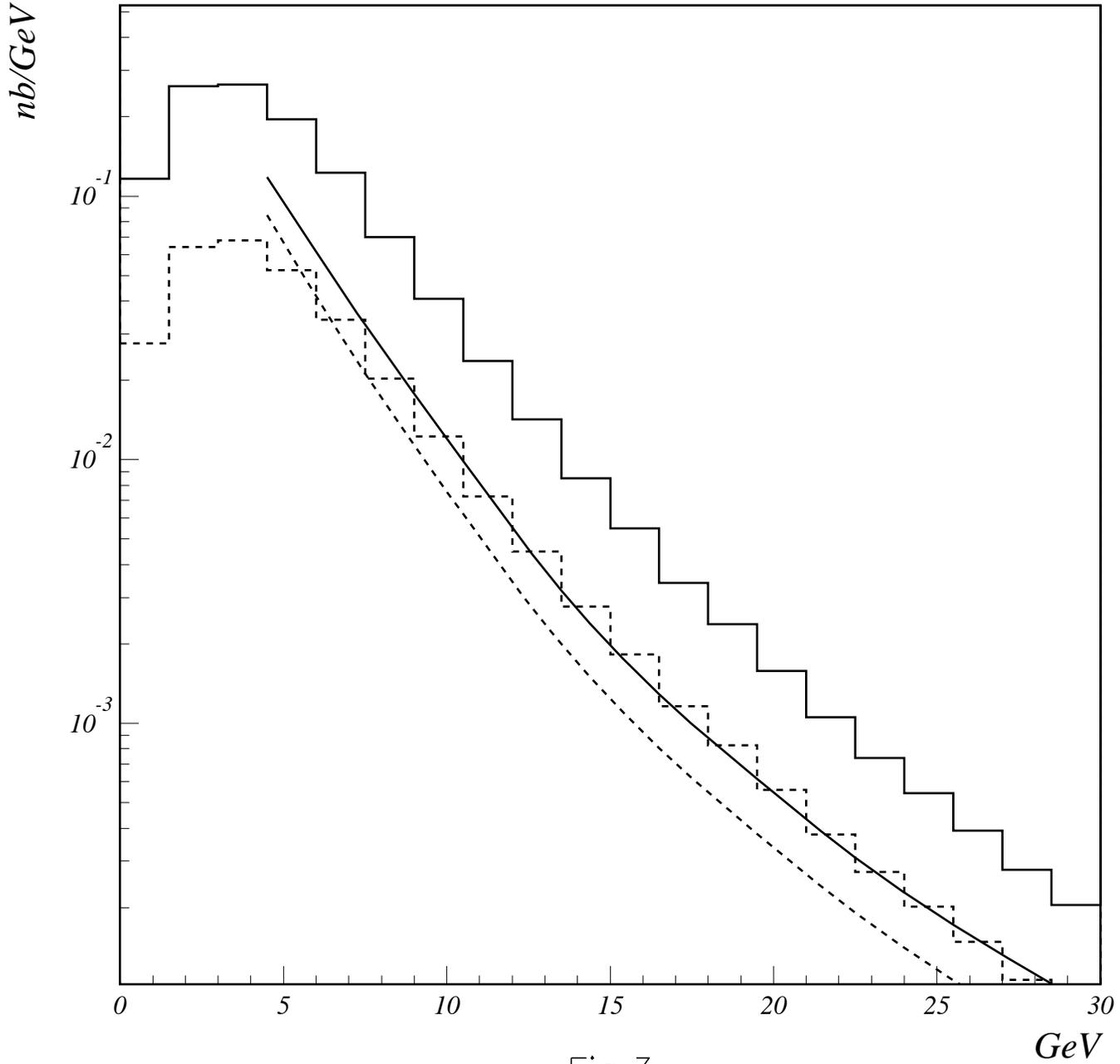
\item[Fig. 3.] The differential cross section
$d\sigma/dp_T$ of the $\Xi_{bc}'(\Xi_{bc}^{(*)})$ production in
$p\bar p$ collisions versus the transverse momentum of the
$\Xi_{bc}'(\Xi_{bc}^{(*)})$-baryon at the hadron interaction energy
1.8 TeV, in comparison with the fragmentation result.

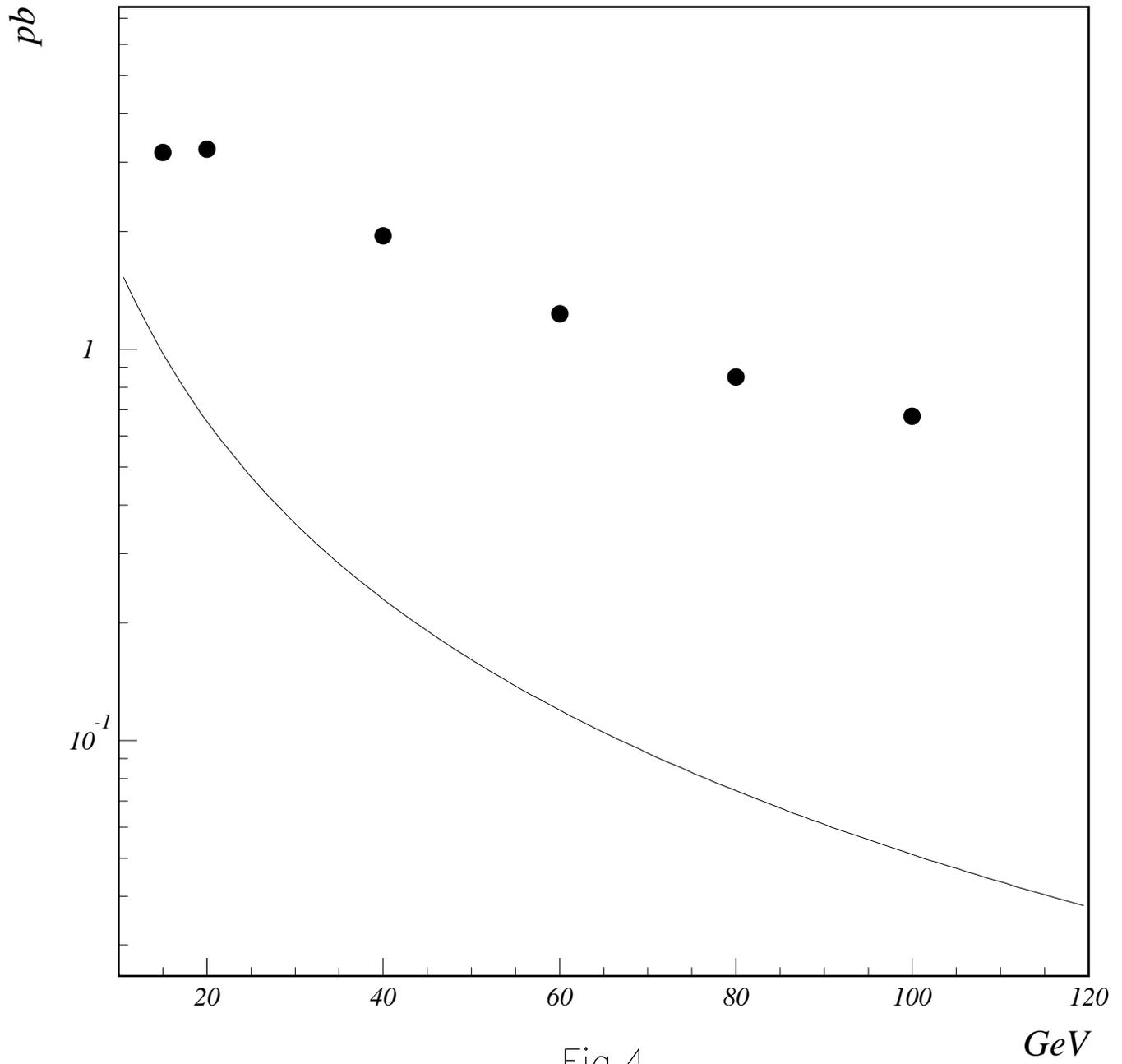
\item[Fig. 4.] The gluonic production cross section of
$\Xi_{cc}^{(*)}$ ($\bullet$) in comparison with
the predictions of the fragmentational mechanism (solid line).

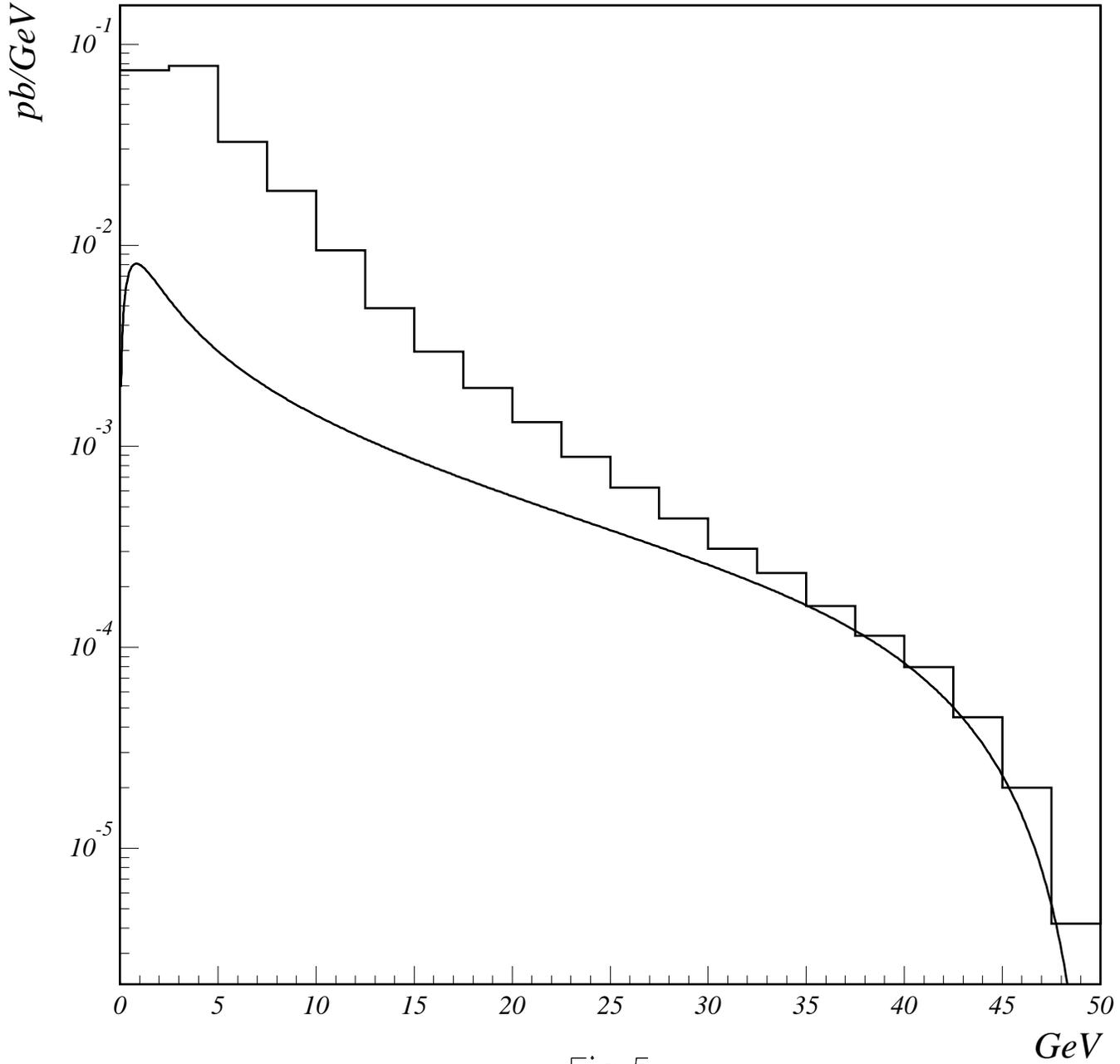
\item[Fig.5.] The distribution over the transverse momentum in the gluonic
production of $\Xi_{cc}^{(*)}$ in comparison with the fragmentation result
at the interaction energy 100 GeV.

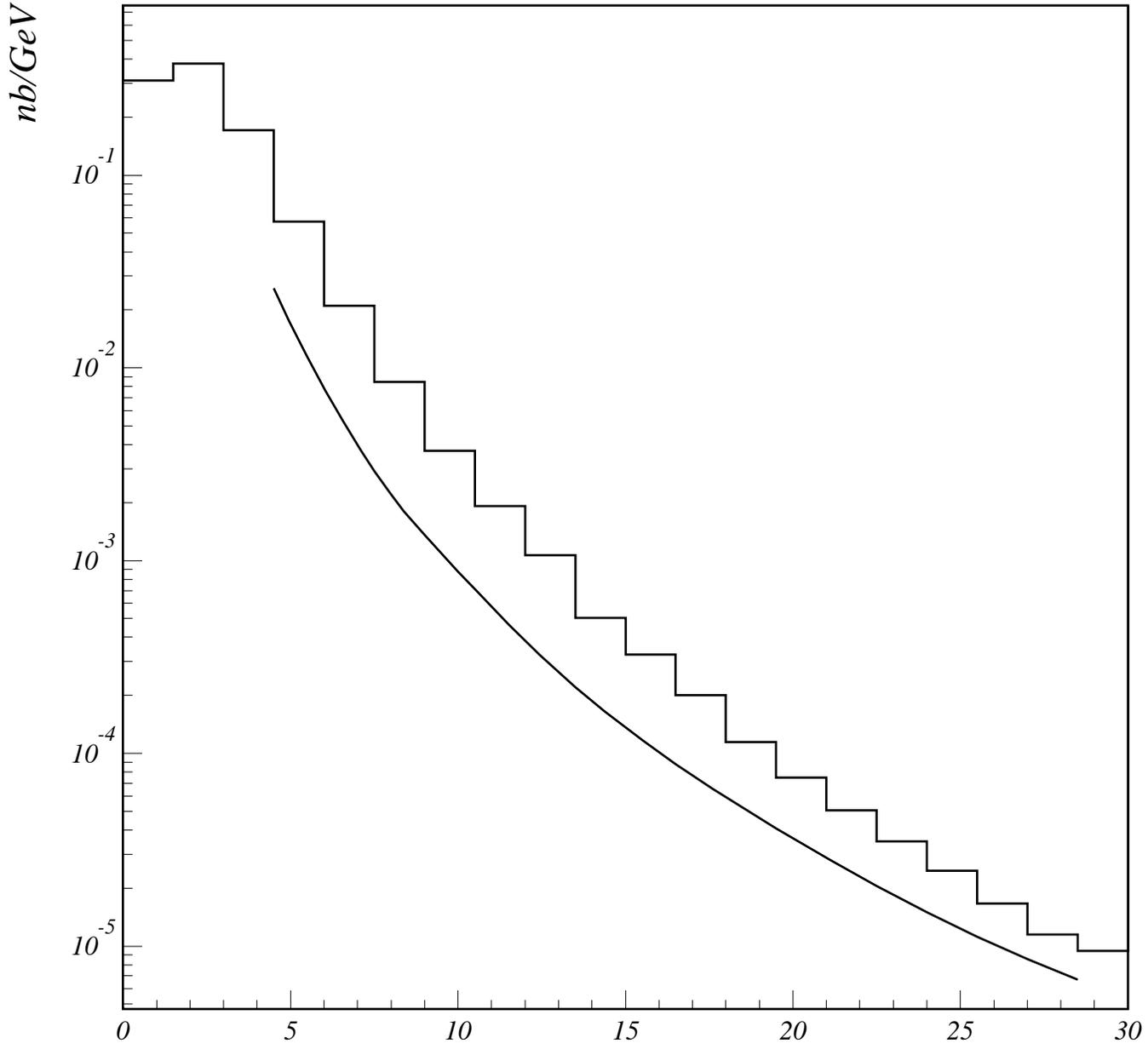
\item[Fig. 6.] The differential cross section
$d\sigma/dp_T$ of the $\Xi_{cc}^{(*)}$ production in
$p\bar p$ collisions versus the transverse momentum of the
$\Xi_{cc}^{(*)}$-baryon at the hadron interaction energy
1.8 TeV, in comparison with the fragmentation result.

\end{itemize}
\hfill {\it Received July 5, 1995}
\end{document}